\documentclass[a4paper,11pt]{article}
\usepackage{amssymb,graphicx,url}

\def\be{\begin{equation}}
\def\ee{\end{equation}}
\def\bea{\begin{eqnarray}}
\def\eea{\end{eqnarray}}

\def\pd{\partial}

\title{Viscous Quark-Gluon Plasma Model Through Fluid QCD Approach\\}

\author{T.P. Djun$^{1,2,*}$ \and B. Soegijono$^{1,3}$ \and T.Mart$^{1,3}$ \and  L.T. Handoko$^{2,4,*}$\\
  \vspace{1mm}\\
  $^{1}$ Graduate Programs in Material Science, University of Indonesia\\
  Kampus UI Salemba, Jakarta 10430, Indonesia \\
  \vspace{1mm}\\
  $^{2}$ Group for Theoretical and Computational Physics\\
  Research Center for Physics, Indonesian Institute of Sciences\\
  Kompleks Puspiptek Serpong, Tangerang 15310, Indonesia\\
  \vspace{1mm}\\
   $^{3}$ Department of Physics, University of Indonesia\\
  Kampus UI Depok, Depok 16424, Indonesia \\
  \vspace{1mm}\\
  $^{4}$ Research Center for Informatics, Indonesian Institute of Sciences\\
  Kompleks LIPI Cisitu, Jl. Cisitu 21/154D, Bandung 40135, Indonesia\\
  \vspace{1mm}\\
  $^{*}$ Email: tpdjun@teori.fisika.lipi.go.id,\\
                  Handoko@teori.fisika.lipi.go.id,\\
                 Laksana.tri.handoko@lipi.go.id\\ 
    }

\date{}

\begin{document}

\maketitle

\begin{abstract}
A Lagrangian density for viscous quark-gluon plasma has been constructed within the fluid-like QCD framework. Gauge symmetry is preserved for all terms inside the Lagrangian, except for the viscous term. The transition mechanism from point particle field to fluid field, and vice versa, is discussed. The energy momentum tensor that is relevant for the gluonic plasma having the nature of fluid bulk of gluon sea is derived within the model. By imposing conservation law in the energy momentum tensor, shear viscosity appears as extractable from the equation. \\
\\
\vspace{2mm}
Keywords : Quark-gluon-plasma, RHIC, Gauge field, Relativistic fluid Lagrangian.
\end{abstract}

\section{Introduction}
\indent As what described by the theory of particle physics, before hadronization came into process there existed a phase of extremely high energy density that comprised of quarks and gluons. Since the last decades, physicists have tried to recreate this phase in laboratories through the ultra-relativistic heavy ion collision experiments. Those experiments reveal a strong indication that a hot dense deconfined phase of free quarks and gluons is found and it constitutes a primordial state of hadronic matter called quark-gluon plasma (QGP). Some key features were investigated, and one of the most interesting properties of QGP is the existence of a very small ratio of shear viscosity over the entropy density. This condition has motivated a number of researches on viscosity of the QGP, and this paper is also written based on such inducement. 

Since our current knowledge about the fundamental hadronic interaction is qualitatively rooted in the non-abelian gauge theory, one can think that the problem can be tackled by means of quantum chromodynamics (QCD). To cope with the complexities that arise due to a large color charge that has to be encountered, lattice gauge theory has thus far become the most preferred calculation tool [1,2]. While from the fact that it behaves like plasma, some other physicists apply the relativistic hydrodynamics approaches to investigate the QGP. Within the hydrodynamics approach, the QGP can be either quark or gluon dominated [3,4,5]. The concept of gluon dominated plasma is motivated by the discoveries of jet quenching in heavy-ion collision at RHIC, indicating the shock waves in the form of March cone [6,7]. While in quark dominated plasma, there is a requisition to have a very small ratio of shear viscosity over entropy to get a good fit of the spectra of transverse momentum or other physical observable to experiment results [8,9,10,11,12,13].\\
\indent This paper is organized as follows. After the introduction given in the first section, the construction of viscous QGP Lagrangian density by using quantum chromodynamics approach is elaborated in the second section. Still in the same section, a kind of phase transition mechanism for gluon field is descibed, where it is explained that from a different point of view the quark gluon field can be considered as a highly energized flow field dominated by its relativistic velocity. The third section is about the construction of the equation of state for shear viscosity. The final part, a discussion, is put at the fourth section.

\section{The Model}
The model, that is described by a complete Lagrangian, consists of a non-viscous part and a viscous term. For the viscous part we adopt the model proposed by Sulaiman et.al. [14], in which the QGP is described as a strongly interacting gluon sea with quarks and anti-quarks inside, and deploys the conventional QCD Lagrangian with SU(3) color gauge symmetry,
\begin{eqnarray}
{\cal L} &=& i \bar{Q} \gamma^\mu \partial_\mu  Q - m_Q \bar{Q} Q - 
{\textstyle \frac{1}{4}} S^a_{\mu\nu} S^{a\mu\nu}  + g_{s } J^{a \mu}_F  U^a_{\mu} .
\label{}
\end{eqnarray}                 
Here  $Q$ is the quark (color) triplet, and $U^a_\mu$ represents the gauge vector field.  $g_s$ is the strong coupling constant, while $J^{a}_{ \mu} = \bar{Q} T^a \gamma_\mu Q$  and  $T^a$'s  belong to the SU(3) Gell-Mann matrices. The field strength tensor is  $S^a_{\mu\nu} \equiv \partial_\mu U^a_\nu - \partial_\nu U^a_\mu 
+ g_F f^{abc} U^b_\mu U^c_\nu$
 , where $f_{abc}$  is the structure constant of the SU(3) group. As to be mentioned that quarks and anti-quarks actually get influenced by the electromagnetic force due to the U(1) field $A^{\mu}$  , but since the corresponding magnitude is very small compared to the case of strong interaction, i.e., $e / g_s = \sqrt{\alpha / \alpha_s} \backsim O (10^{-1})$  , it may practically be neglected. 

Following the original model [14,15], the gluon fluid is constructed to have a particular form in term of the relativistic velocity as, 
\begin{equation}
U^a_\mu = (U^a_0 , \textbf{U}^a) = u^a_\mu \phi 
\end{equation}
where $u^a_\mu = \gamma_{V^a}  (1, 
\textbf{v}^a)$ and $\gamma_{V^a}  = (1-\vert\textbf{v}^a\vert^2)^{-1/2}$.  To keep the correct dimension, we define $\phi$ as a dimension  one scalar field, and it represents the field distribution. Within this formulation, the equation of motion (EOM) for single gluon field derived from the Lagrangian above takes the form of 
\begin{equation}
\frac{\partial}{\partial t} (\gamma_{V^a} \mathbf{v}^a \phi) + \nabla 
(\gamma_{V^a} \phi) = -g_s \oint d\mathbf{x} (J^a_{0} + F^a_0).
\label{eq:eq-of-motion}
\end{equation}

\noindent Note that $J^a_0$ is a covariant current of gluon field, and $F^a_0$ is an auxiliary function. Equation (3) can be considered as a general relativistic fluid equation, since at the non-relativistic limit the equation reduces to the classical Euler equation. As a consequence, this fact leads to the interpretation that a gluon particle $U^a_\mu$  at a certain scale might behave as a fluid field. One can then consider this as a kind of  "phase transition",\\
\\
$\underbrace{hadronic \; \; state}_{\epsilon^a_\mu}  \longleftrightarrow \underbrace{QGP \; state}_{u^a_\mu} .$
\\
As the gluon field behaves like a point particle, it is in a stable hadronic state, and characterized by its polarization vector $\xi_\mu$ as usually written in the form of $U^a_\mu = \xi^a_\mu \phi$.  However, as it is near to the hadronization, like the hot QGP, it behaves like a flow field with properties dominated by its relativistic velocity.
Now, we move forward to discuss the viscous term in the model. The Lagrangian of the viscous term is derived from the viscous energy momentum tensor. In this study we adopt the standard viscous energy momentum tensor that is used at the non-equilibrium quantum field theory or relativistic hydrodynamics [16,17,18]. The relevant formulation reads

\begin{eqnarray}
\mathcal{T}^{\mu\nu}_{vis} &=& - c \eta T^a (\partial^\nu U^\mu + \partial^\mu U^\nu - 
  U^\nu U^l \partial_l U^\mu - U^\mu U^l \partial_l U^\nu) \nonumber\\
  &&- c (\zeta - {\textstyle \frac{2}{3}} 
  \eta )  \partial_l U^l (g^{\mu\nu} - U^\mu U^\nu).
\end{eqnarray}
By taking the natural unit $c = 1$ , and assuming that the bulk viscosity $\zeta$ is negligible in the gluonic plasma, one gets
\begin{eqnarray}
\mathcal{T}^{\mu\nu}_{vis} &=& - \eta T^a (\partial^\nu U^\mu + \partial^\mu U^\nu - 
  U^\nu U^l \partial_l U^\mu - \nonumber\\
  &&- U^\mu U^l \partial_l U^\nu) +  {\textstyle \frac{2}{3}} 
  \eta   \partial_l U^l (g^{\mu\nu} - U^\mu U^\nu).
\end{eqnarray}
For the reverse calculation, we make use of the energy momentum tensor equation $\mathcal{T}^{\mu\nu} = \frac{2}{\sqrt{-g}} \frac{\delta \mathcal{L}}{\delta g_{\mu\nu}} $ . \\
By using the identity  $\sqrt{-g} \; \delta g_{\mu\nu} = -\frac{1}{2} g_{\mu\nu} \; \delta\sqrt{-g}$, one can get  $\mathcal{L} = -\frac{1}{4} \mathcal{T}^{\mu\nu} \delta g_{\mu\nu} \;\sqrt{-g}$ . 

From the viscous energy momentum tensor, the reverse calculation process can be performed at a term-by-term basis. For example, for the first term\\
\begin{eqnarray}
\delta \mathcal{L}_I &=& -\frac{1}{4} \eta T^a (-\partial_\nu U^a_\mu) g^{\mu\nu} \delta \sqrt{-g}\nonumber\\ 
\mathcal{L}_I &=& -\frac{1}{4} \eta T^a (-\partial_\nu U^{a\mu} \sqrt{-g}. \nonumber
\end{eqnarray}

\noindent Then, by summing back all the terms, $\mathcal{L}_I + \mathcal{L}_II + ......+ \mathcal{L}_VI$ , one gets

\begin{eqnarray}
{\cal L}_{vis} &=& \frac{1}{4} \eta T^a ( \partial_\nu U^{a \nu} +    \partial^\nu U^a_\nu -  U^{a \mu} U^{b l} \partial_l U^b_\mu - U^{a \nu} U^{b l} \partial_l U^b_\nu) \nonumber \\
&&-  \frac{2}{3} \eta T^a U \partial_l U^{a l}  + \frac{1}{6} \eta T^a U \partial_l U^{a l} \nonumber\\
&=& \frac{1}{4} \eta T^a ( \partial_\mu U^{a \mu} +    \partial^\mu U^a_\mu -  2 U^{a \mu} U^{b l} \partial_l U^b_\mu) - \frac{1}{2} \eta T^a \partial_l U^{a l} \nonumber\\
&=& \frac{1}{4} \eta T^a ( \partial_\mu U^{a \mu} +    \partial^\mu U^a_\mu - 2 U^{a \mu} U^{b l} \partial_l U^b_\mu - 2  \partial_l U^{a l} ). 
\label{eq:Lagrangian}
\end{eqnarray}
The above simplification is made under the assumption that the symmetry between $\mu$  and $\nu$ is valid. Further, since $\partial_\mu U^{a \mu} = \partial^\mu U^a_\mu$ , and due to the dummy index  $\mu$ and $l$ , it brings to the condition of $ \partial_\mu U^{a \mu} +    \partial^\mu U^a_\mu - 2 \partial_l U^{a l} = 0$ . The final result for the Lagrangian density of the viscous term reads
\begin{equation}
\mathcal{L}_{vis} = -\frac{1}{2} \eta T^a U^{a \mu} U^{b l} \partial_l U^b_\mu,
\end{equation}
whereas, the total Lagrangian becomes
\begin{eqnarray}
\cal{L} &=& i \bar{Q} \gamma^\mu \partial_\mu  Q - m_Q \bar{Q} Q - 
{\textstyle \frac{1}{4}} S^a_{\mu\nu} S^{a\mu\nu}  + g_{s } J^a_\mu  U^a_{\mu}\nonumber \\
&&- \frac{1}{2} \eta T^a U^{a\mu} U^{bl} \pd_l U^b_\mu .
\end{eqnarray}

\section{Constructing The Shear Viscosity}

In formulating the shear viscosity, the steps will be started from obtaining the complete energy momentum tensor $\mathcal{T}^{\mu\nu} = \mathcal{T}^{\mu\nu}_{non-vis} + \mathcal{T}^{\mu\nu}_{vis}$ . 
Note that  $ \mathcal{T}^{\mu\nu}_{vis}$ could be obtained from Eq. (1) by utilizing $\mathcal{T}^{\mu\nu} = \frac{2}{\sqrt{-g}} \frac{\delta \mathcal{L}}{\delta g_{\mu\nu}} $  .  For the present discussion, we tend to assume that the matter concerned is gluon dominated QGP, so that in this case the quark and anti-quark terms are neglected due to their minor contributions to the system.  The result is 
\begin{equation}
\mathcal{T}^{\mu\nu}_{non-vis} = S^{a \mu\nu} S^{a \rho}_\nu - g^{\mu\nu} \mathcal{L} + 2 g_s J^{a \mu} U^{a \nu}.
\end{equation}
Then, for the sake of simplicity, the gluon fields are assumed to be similar for all color states, i.e.  $u^a_\mu = U_\mu$  for $a = 1, 2, 3,......, 8$    ,  which yields
\begin{eqnarray}
\mathcal{T}^{\mu\nu}_{vis} &=& [8 g_F f_Q m_Q \phi + g^2_F f^2_g \phi^4] u^\mu u^\nu  - [4 g_F f_Q m_Q \phi - {\textstyle \frac{1}{4}} g^2_F f^2_g \phi^4] g^{\mu\nu}.
\label{eq:energy-momentum-tensor}
\end{eqnarray}

\noindent The total energy momentum tensor becomes
\begin{eqnarray}
\mathcal{T}^{\mu\nu} &=& \mathcal{T}^{\mu\nu}_{non-vis} + \mathcal{T}^{\mu\nu}_{vis} \nonumber\\
&=& [8 g_F f_Q m_Q \phi + g^2_F f^2_g \phi^4] u^\mu u^\nu - [4 g_F f_Q m_Q \phi - {\textstyle \frac{1}{4}} g^2_F f^2_g \phi^4] g^{\mu\nu}\nonumber\\
&&- \eta T (\partial^\nu U^\mu + \partial^\mu U^\nu - U^\nu U^l \partial_l U^\mu- U^\mu U^l \partial_l U^\nu) \nonumber \\
&&+  {\textstyle \frac{2}{3}} \eta \partial_l U^l (g^{\mu\nu} - U^\mu U^\nu)
\end{eqnarray}  
The shear viscous components of the energy momentum tensor are assumed to be symmetric, i.e., $t^{\mu\nu} = t^{\nu\mu}$ . Then, it turns to a simpler form, 
\begin{eqnarray}
\mathcal{T}^{\mu\nu}_{total} &=& \mathcal{T}^{\mu\nu}_{non-vis} + \mathcal{T}^{\mu\nu}_{vis} \nonumber\\
&=& [8 g_F f_Q m_Q \phi + g^2_F f^2_g \phi^4] u^\mu u^\nu - [4 g_F f_Q m_Q \phi - {\textstyle \frac{1}{4}} g^2_F f^2_g \phi^4] g^{\mu\nu}\nonumber\\
&& - \eta T (\frac{14}{3}\partial^\mu u^\nu \phi - \frac{16}{3}\partial^\mu u^\nu \phi^4 )
\end{eqnarray}  
By utilizing one of the most powerful tools in theoretical physics, i.e. the conservation law, a number of observables can be extracted from a system or an equation. Here, we apply a covariant derivative to the energy momentum tensor, $\nabla_\mu \mathcal{T}^{\mu\nu} = \frac{1}{\sqrt{-g}} \pd_\mu \mathcal{T}^{\mu\nu} + \Gamma^{\nu}_{\sigma \mu} \mathcal{T}^{\mu\sigma} = 0$.\\
For the sake of compactness, let us set $g_s f_Q m_Q = \alpha$ and $g^2_s f^2_g = \beta$. Then
\begin{eqnarray}
\nabla_\mu \mathcal{T}^{\mu\nu}_1 &=&   \frac{1}{\sqrt{-g}} \pd_\mu  [8 g_s f_Q m_Q \phi + g^2_s f^2_g \phi^4] u^\mu u^\nu + \Gamma^{\nu}_{\sigma \mu}  [8 g_s f_Q m_Q \phi + g^2_s f^2_g \phi^4] u^\mu u^\sigma. \nonumber\\
&=& \frac{1}{\sqrt{-g}} [8 \alpha (\pd_\mu \phi) + \beta (\pd_\mu \phi^4)]u^\mu u^\nu +  \frac{1}{\sqrt{-g}} [8 \alpha \phi + \beta \phi^4](\pd_\mu u^\mu) u^\nu \nonumber\\
&&+ \frac{1}{\sqrt{-g}} [8 \alpha \phi + \beta \phi^4] u^\mu (\pd_\mu u^\nu) + \Gamma^\nu_{\sigma \mu} ( [8 \alpha \phi + \beta \phi^4] u^\mu u^\nu) ,
\end{eqnarray}
and
\begin{eqnarray}
\nabla_\mu \mathcal{T}^{\mu\nu}_2 &=& g^{\mu\nu}  \frac{1}{\sqrt{-g}} [4\alpha (\pd_\mu \phi) - \frac{1}{4} \beta (\pd_\mu \phi^4)] + \Gamma^\nu_{\nu \mu} [(4\alpha \phi - \frac{1}{4} \beta \phi^4) g^{\mu\nu}].
\end{eqnarray}
\begin{eqnarray}
\nabla_\mu \mathcal{T}^{\mu\nu}_3 &=&   \frac{14}{3 \sqrt{-g}} \eta  [2 (\pd_\mu \pd^\mu u^\nu) \phi + \pd^\mu u^\nu (\pd_\mu \phi)] + \Gamma^\nu_{\sigma \mu} (\frac{14}{3} \eta  \pd^\mu u^\sigma \phi) \nonumber\\
&=&  \frac{14}{3 \sqrt{-g}} \eta  [2 (\pd_\mu \pd^\mu) u^\mu \phi +  g^{\mu\nu}  \pd^\mu u_\mu (\pd_\mu \phi)] + \Gamma^{\nu \mu}_{ \mu} g_{\mu\sigma}  (\frac{14}{3} \eta  g^{\mu \sigma}  \pd^\mu u_\sigma \phi)\nonumber\\
&=&  \frac{14}{3 \sqrt{-g}} \eta  [2 (\pd_\mu \pd^\mu) u^\mu \phi +  g^{\mu\nu}  \pd^\mu u_\mu (\pd_\mu \phi)] + 4 \Gamma^{\nu \mu}_{ \mu}   (\frac{14}{3} \eta   \pd^\mu u_\mu \phi),
\end{eqnarray}
as well as
\begin{eqnarray}
\nabla_\mu \mathcal{T}^{\mu\nu}_4 &=&   \frac{16}{3 \sqrt{-g}} \eta T [2 (\pd_\mu \pd^\mu u^\nu) \phi^4 + \pd^\mu u^\nu (\pd_\mu \phi^4)] + \Gamma^\nu_{\sigma \mu} (\frac{16}{3} \eta  \pd^\mu u^\sigma \phi^4)\nonumber\\
&=&  \frac{16}{3 \sqrt{-g}} \eta  [2 (\pd_\mu \pd^\mu u^\nu) \phi^4 + g^{\mu\nu}  \pd^\mu u_\mu (\pd_\mu \phi^4)] + 4 \Gamma^{\nu\mu}_{\mu} (\frac{16}{3} \eta \pd^\mu u_\mu \phi^4).
\end{eqnarray}
\\
By adding up all terms,\\
\\
$\nabla_\mu  \mathcal{T}^{\mu\nu}_{total} = \nabla_\mu [ \mathcal{T}^{\mu\nu}_1 + \nabla_\mu  \mathcal{T}^{\mu\nu}_2 + \nabla_\mu \mathcal{T}^{\mu\nu}_3 + \nabla_\mu  \mathcal{T}^{\mu\nu}_4 ] = 0$,\\
one get,
\begin{eqnarray}
\nabla_{\mu}  \mathcal{T}^{\mu\nu}_{total} &=& \nabla_\mu [\mathcal{T}^{\mu\nu}_{1} + \mathcal{T}^{\mu\nu}_{2} +  \mathcal{T}^{\mu\nu}_{3} +  \mathcal{T}^{\mu\nu}_{4} ] = 0 \nonumber
\end{eqnarray}
\begin{eqnarray}
\nabla_\mu  \mathcal{T}^{\mu\nu}_{total} &=& [\frac{1}{\sqrt{-g}}  (8\alpha \phi + \beta \phi^4) u^\nu + \frac{14}{3\sqrt{-g}} \pd_\mu \phi g^{\mu\nu}  4\Gamma^{\nu\mu}_\mu \frac{14}{3} \eta  \phi - \frac{16}{3\sqrt{-g}} \eta  (\pd_\mu \phi^4) g^{\mu\nu} \nonumber\\
&-&  4\Gamma^{\nu\mu}_\mu  \frac{16}{3\sqrt{-g}} \eta  \phi^4 ] \pd^\mu u_\mu + \frac{1}{\sqrt{-g}} (8 \alpha \phi + \beta \phi^4) u^\mu (\pd_\mu u^\nu) = 0
\end{eqnarray}
Currently, this equation can be considered as an equation of state that is still wide open for various elaborations.

\section{Discussion}

The conservation equation of the energy momentum tensor has a very rich structure and content. By imposing certain conditions, at some points the conservation equation can be simplified and some observables can be directly derived. Or, through an indirect approach, the conservation equation can be fed further to other theory for obtaining some other observables, like in the case of utilizing linear response theory and Kubo formula for extracting the shear and bulk viscosities, etc. 
       
One of simpler ways to look for shear viscosity is by taking the conservation equation for energy density  $\mathcal{T}^{0 \nu}$ , where the energy density is taken from the first row component of the matrix of energy momentum tensor. 
Simplify further the conservation equation by taking the local inertial frame, $\pd_\nu \mathcal{T}^{0 \nu} = 0$  , one gets a continuity equation for energy density,
\begin{eqnarray}
\pd_\nu \mathcal{T}^{0 \nu} = \pd_0 \mathcal{T}^{00} + \pd_1 \mathcal{T}^{01} + \pd_2 \mathcal{T}^{02} + \pd_3 \mathcal{T}^{03} = 0  \nonumber     
\end{eqnarray}
Some elaborations on the equation is expected to bring a closer stand to the shear viscosity coefficient.

Several works on utilizing the non-viscous energy momentum tensor have been done elsewhere [19,20]. The observables are derived and calculated following the logic of physics theories and procedures, and the preciseness of the results is waiting for comparison with other?s works in the same field. The addition of viscous term in the energy momentum tensor is expected to act as a fine tuner variable for the investigated observables.

\section{Acknowledgment}
TPD thanks the Group for Theoretical and Computational Physics, Research Center for Physics, Indonesian Institute of Sciences (LIPI) for warm hospitality during the work. This work is funded by Riset Kompetitif LIPI in fiscal year 2013 under Contract no. 11.04/SK/KPPI/II/2013. 

\section{References}

\begin{enumerate}
\item S.Gottlieb, {\itshape J. Phys. Conf. Ser.} 78, 012023 (2007).
\item P. Petreczky, {\itshape Europ. Phys. J. Special Topics} 155, 1951 (2008).
\item I. Bouras, E. Molmar, H. Niemi, Z. Xu, A. El, O. Fochler, C. Greiner, D. Rischke, {\itshape Phys. Rev. Lett.} 103, 032301 (2009).
\item I. Bouras, A. El, O. Fochler, J. Uphoff, Z. Xu, C. Greiner, {\itshape arXiv:0906.2675v1 [hep-ph].}
\item P. Romatschke, {\itshape Int. J. Mod. Phys. E} 19, 1(2010).
\item J. Adams, et.al. (STAR Collaboration), {\itshape Phys. Rev. Lett. }91, 172302 (2003).
\item A. Adare, et.al. (PHENIX Collaboration), {\itshape Phys. Rev. Lett.} 101, 232301 (2008).
\item	D. Teaney, J. Lauret, E. V. Shuryak, {\itshape Phys. Rev. Lett.}  86, 4783 (2001).
\item	P. Huovinen, P. F. Kolb, U. W. Heinz, P. V. Ruuskanen, S. A. Voloshin, {\itshape Phys. Lett. B,} 503, 58 (2001).
\item	P. F. Kolb, U. W. Heinz, P. Huovinen, K. J. Eskola, K. Tuominen, {\itshape Nucl. Phys. A,} 696, 197 (2001).
\item	P. F. Kolb, R. Rapp, {\itshape Phys. Rev. C,} 67, 044903 (2003).
\item	T. Hirano, K. Tsuda, {\itshape Phys. Rev. C,} 66, 054905 (2002).
\item	R. Baier, P. Romatschke, {\itshape Eur. Phys. J. C} 51, 677 (2007).
\item	A. Sulaiman, A. Fajarudin, T. P. Djun and L. T. Handoko, {\itshape International Journal of Modern Physics A} 18-19, 3630-3637 (2009).
\item	T. P. Djun, L. T. Handoko, in {\itshape Proceeding of the Conference in Honour of Murray Gell-Mann's 80th Birthday : Quantum Mechanics, Elemantary particles, Quantum Cosmology and Complexity} (2011) pp.419-425. 
\item	L. D. Landau and E. M. Lifshitz, Course of Theoretical Physics Volune 6, {\itshape ?Fluid Mechanics?.}, Reed Educational and Professional Publishing Ltd. (1987). 
\item	E. A. Calzetta, B. B. Hu, {\itshape Nonequilibrium Quantum Field Theory.} Cambridge University Press (2008)
\item	R. L. Liboff, {\itshape Kinetic Theory, Classical, Quantum, and Relativistic Descriptions.} Springer-Verlag New York, Inc. (2003)
\item	C. S. Nugroho, A. O. Latief, T. P. Djun and L. T. Handoko, {\itshape Gravitation and Cosmology} 18, 32 (2012). 
\item	A. Sulaiman, T. P. Djun, L. T. Handoko, {\itshape J. Theor. Comput. Stud. } 5, 0401 (2006).
\end{enumerate}

\end{document}